\documentclass[%
 reprint,
 amsmath,amssymb,
 aps,
]{revtex4-2}

\usepackage{svg}
\usepackage{siunitx}

\usepackage{graphicx}
\usepackage{dcolumn}

\begin{document}

\preprint{APS/123-QED}

\title{Picosecond acoustic dynamics\\in stimulated Brillouin scattering}

\author{Johannes Piotrowski}
\email{jpiotrowski@ethz.ch}
\affiliation{ETH Zürich, Hönggerbergring 64, 8093 Zürich, Switzerland}
\affiliation{Macquarie University Research Centre in  Quantum Engineering (MQCQE), MQ Photonics Research Centre, Department of Physics and Astronomy, Macquarie University, NSW 2109, Australia.}

\author{Miko\l{}aj K. Schmidt}
\affiliation{Macquarie University Research Centre in  Quantum Engineering (MQCQE), MQ Photonics Research Centre, Department of Physics and Astronomy, Macquarie University, NSW 2109, Australia.}
\author{Birgit Stiller}
\affiliation{Max Planck Institute for the Science of Light, Staudtstraße 2, 91058 Erlangen, Germany}
\affiliation{Institute of Photonics and Optical Science (IPOS), School of Physics, University of Sydney, NSW 2006, Australia}
\author{Christopher G. Poulton}
\affiliation{School of Mathematical and Physical Sciences, University of Technology Sydney, NSW 2007, Australia.}
\author{Michael J. Steel}
\affiliation{Macquarie University Research Centre in  Quantum Engineering (MQCQE), MQ Photonics Research Centre, Department of Physics and Astronomy, Macquarie University, NSW 2109, Australia.}




\begin{abstract}
Recent experiments demonstrating storage of optical pulses in acoustic phonons based on stimulated Brillouin scattering raise a number of questions about the spectral and temporal capacities of such protocols and the limitations of the theoretical frameworks routinely used to describe them. In this work, we consider the dynamics of photon-phonon scattering induced by optical pulses with temporal widths comparable to the period of acoustic oscillations. We revisit the widely adopted classical formalism of coupled modes and demonstrate its breakdown. We propose a simple extension to generalise the formulation and find potentially measurable consequences in the dynamics of Brillouin experiments involving ultra-short pulses.
\end{abstract}

\maketitle

Stimulated Brillouin Scattering (SBS), the coherent scattering of light waves and acoustic vibrations, has in recent years been reinvigorated~\cite{Garmire2018} as a powerful tool for shaping optical waves. Confined in optoacoustic waveguides, it can be used for a wide range of optical processing applications~\cite{Eggleton2019}: narrow-linewidth filtering~\cite{Vidal2007,Marpaung2015}, controlling coherent cross-talk between waveguides~\cite{Shin2015}, signal integration and derivation \cite{Sangiustina2013}, optoacoustic delay~\cite{Song2009,Antman2012} and storage~\cite{Zhu2007,Preussler2009,Merklein2017,Stiller2019} as well as optical signal demodulation~\cite{Giacoumidis2018}. Many of these experiments require harnessing Brillouin effects for short optical pulses --- that is, for pulses with durations shorter than the acoustic lifetime in the material, typically \SI{1}-\SI{10}{\nano\second}. Short pulses are used either to broaden the Brillouin gain spectrum for broadband signal processing or to ensure that the pulses fit entirely into a waveguide of finite length. The application of SBS to further extremes --- where the optical pulses are in the picosecond regime --- although difficult due to the short interaction time~\cite{Shapiro1967}, has nevertheless been demonstrated in chalcogenide waveguides~\cite{Jaksch2017art}. However, the conventional formalism of Brillouin scattering in waveguides is based on the assumption that the pulses remain long, and it is currently unclear where this formalism remains applicable and where it breaks down. An analysis capable of accurately describing SBS in waveguides in the picosecond regime is needed for the full understanding of short-pulse Brillouin experiments.

In this paper, we theoretically explore SBS in the regime where the optical pulses are shorter than the acoustic period. The usual tool for describing Brillouin scattering is the coupled mode formalism~\cite{Rakich2010,Wolff2015waveguides} (for clarity referred to here as the {\em conventional formalism}). We examine the assumptions and limitations of this model for short pulses, and show where it remains applicable. 
Revisiting the underlying approximations --- in particular the Slowly-Varying-Envelope Approximation (SVEA) and the Rotating-Wave Approximation (RWA)~\cite{Siegman1986,Sazonov2017} --- we show that it is necessary to retain the second order derivative in time of the acoustic field, while neglecting the second order derivative in space remains appropriate. We apply the resulting model to the protocol of acoustic memory storage. We find that the abandoning of the underlying approximations leads not only to a more accurate model, but predicts physical effects that dominate the dynamics of SBS in the short-pulse regime, observable for any Brillouin system once pulses are sufficiently short.
We note that related issues have been explored for opto-elastic interactions in liquids where the retention of a higher order derivative has also been found appropriate~\cite{Yuan2019,Velchev1999}.

\begin{figure}[!ht]
    \includegraphics[width=.95\linewidth]{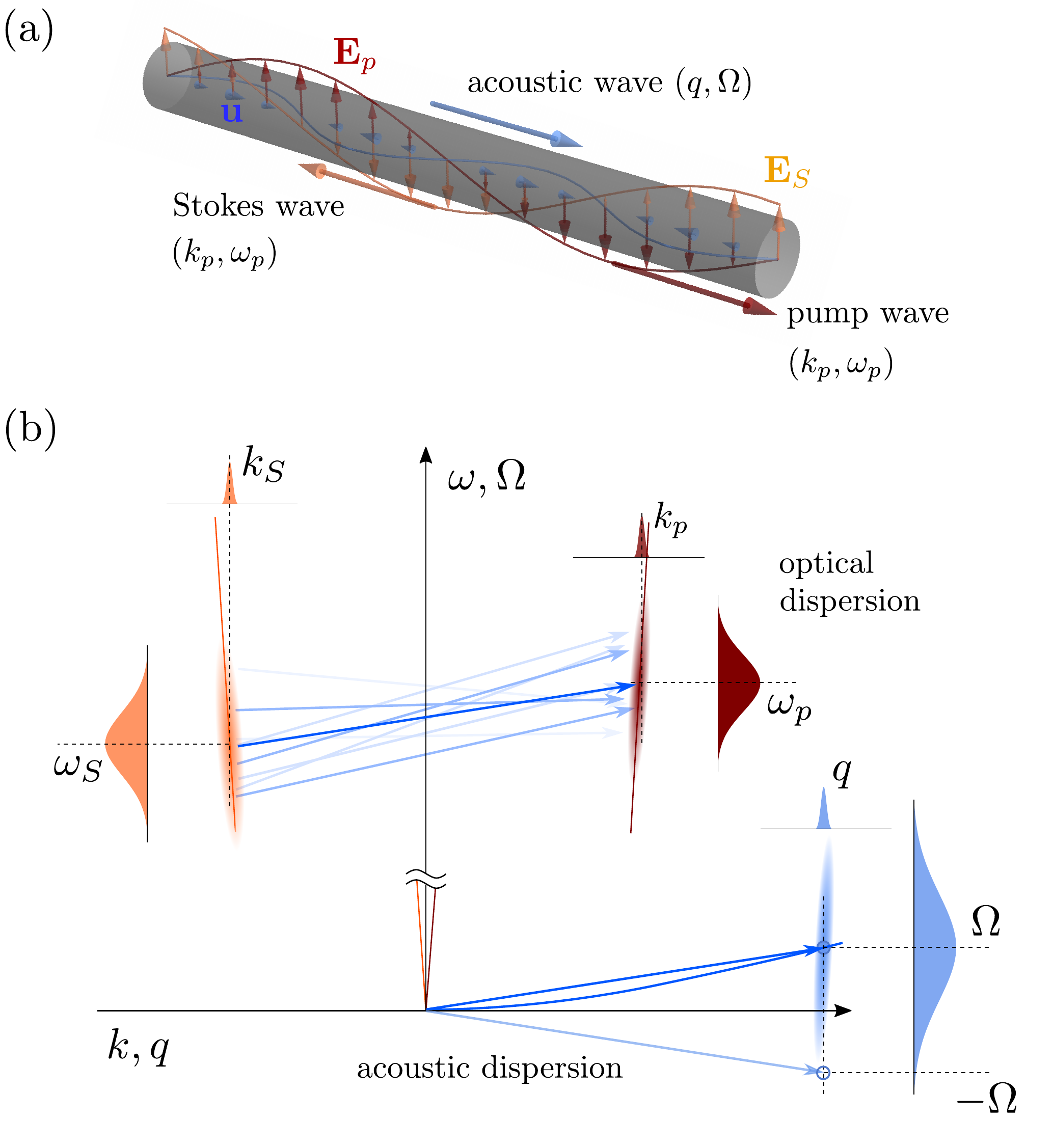}
    \caption{(a): Pump (red), signal (orange) and acoustic (blue) modes of the optical and acoustic fields respectively interact within a waveguide as described in the text. (b): A sketch of the phase and energy matching conditions during SBS interactions on the optical dispersion branches with the acoustic wave as a connecting vector reveals the difference in scales between wavenumbers and frequencies for wave pulses.}
    \label{fig:setup}
\end{figure}

The widely adopted classical formalism for SBS in waveguides~\cite{Rakich2010,Wolff2015waveguides} describes the spatial and temporal evolutions of three coupled complex envelope functions associated with the amplitude of the pump ($a_\mathrm{p}$), signal ($a_\mathrm{s}$) and acoustic ($b$) waves:
\begin{align} \label{eq:pump}
\partial_{z} a_{\mathrm{p}}+\frac{1}{v_{\mathrm{p}}} \partial_{t} a_{\mathrm{p}} &=-\frac{i \omega_{\mathrm{p}} }{\mathcal{P}_{\mathrm{p}}} \mathcal{Q}_{\mathrm{p}} a_{\mathrm{s}} b^{*}\;,\\
 \label{eq:Stokes}
\partial_{z} a_{\mathrm{s}}+\frac{1}{-v_{\mathrm{s}}} \partial_{t} a_{\mathrm{s}} &=-\frac{i \omega_{\mathrm{s}} }{\mathcal{P}_{\mathrm{s}}} \mathcal{Q}_{\mathrm{s}} a_{\mathrm{p}} b \;,\\
 \label{eq:acoustic}
\partial_z b +  \frac{1}{v_\mathrm{a}}\partial_t  b   + \alpha b
    &= -\frac{i\Omega}{\mathcal{P}_\mathrm{o}}\mathcal{Q}_\mathrm{a} a_{\mathrm{p}}a_{\mathrm{s}}^*\;.
\end{align} 
These envelopes, represented schematically in Fig.~\ref{fig:setup}(a), are related to the electric and elastic displacement fields of the optical and acoustic modes, and their respective transverse mode profiles (marked as tilded vector fields), as $\mathbf{E}_{p/s}(\mathbf{r},t) = a_p(z,t) \tilde{\mathbf{e}}_{p/s}(\mathbf{r}_\perp) \exp[i(k_{p/s}z - \omega_{p/s}t)] + \mathrm{c.c.}$ and $\mathbf{U}(\mathbf{r},t) = b(z,t) \tilde{\mathbf{u}}(\mathbf{r}_\perp) \exp[i(qz - \Omega t)] + \mathrm{c.c.}$. Temporal and spatial evolution of the pump/signal/acoustic waves is governed by frequencies $\omega_\mathrm{p}$/$\omega_\mathrm{s}$/$\Omega$
and wavenumbers $k_\mathrm{p}$/$k_\mathrm{s}$/$q$ respectively ($k_\mathrm{s}<0$), which are connected
by the group velocities $v_\mathrm{p}$/$v_\mathrm{s}$/$v_\mathrm{a}$ (all positive consistent with Eqs~\ref{eq:pump}). The quantities $P_\mathrm{p/s/a}$ and $Q_\mathrm{p/s/a}$ describe the mode powers for normalisation and coupling constants, respectively, and $\alpha$ is the linear loss of the acoustic wave.
On backwards SBS resonance, these parameters are related by the phase- and energy-matching relations $k_\mathrm{p}-k_\mathrm{s}=q$ and $\omega_\mathrm{p}-\omega_\mathrm{s} = \Omega$ (sketched in Fig.~\ref{fig:setup}~(b)) and conservation of energy also requires~\cite{Wolff2015waveguides} $\mathcal{Q}_\mathrm{a} = (\mathcal{Q}_\mathrm{p} + \mathcal{Q}_\mathrm{s})/2$.

The above formulation relies on two key physical approximations: the slowly-varying envelope approximation (SVEA) and the rotating wave approximation (RWA), each of them in a temporal and spatial dimension. Limitations to these approximations are given by the dynamics of the acoustic wave at short time scales. We therefore review the derivation of \eqref{eq:acoustic} in a limit where the frequency and wavenumber spectra of the optical and acoustic waves are as shown in Fig.~\ref{fig:setup}(b), i.e. they are well-defined in $q$, but their frequency spectra have widths comparable to $\Omega$. This is intuitively justified by considering the physical parameters of a picosecond Brillouin setup. Counter-propagating pump and signal pulses of width $\tau$ will interact within domains spanning $(v_\mathrm{p}+v_\mathrm{s}) \tau/2$ and $\tau/2$ in space and time, respectively, which define the widths of the wavenumber $\Delta q \sim [(v_\mathrm{p}+v_\mathrm{s}) \tau]^{-1}$ and frequency $\Delta \Omega\sim \tau^{-1}$ spectra of the optical force that induces acoustic excitations. The spatial RWA and SVEA are valid if the wavenumber spectrum  is narrow: $\Delta q \ll q$. Substituting the above approximate expression for $\Delta q$, and using $q\approx \Omega/v_a$, we find that this condition breaks down for $\tau \sim v_a/(v_\mathrm{p}+v_\mathrm{s})/\Omega\sim 10^{-15}$~s, where we have taken $v_a = \SI{2595}{\meter\per\second}$, $v_p=v_s=\SI{8.9E7}{\meter\per\second}$, and $\Omega/2\pi = \SI{10}{\giga\hertz}$. On the other hand, the limits of the temporal SVEA are reached when $\Delta \Omega \sim \Omega$, or for $\tau \sim \Omega^{-1} \sim 10^{-10}$~s pulses. Therefore, Fig.~\ref{fig:setup}(b) accurately represents the spectrum of the dispersion of the optical force induced by picosecond optical pulses. It also emphasises the possibility of coupling to the backward-propagating acoustic modes, which forces us to embrace a more complete picture of the acoustic pulse dynamics.

Acoustic waves are described by the elastic displacement field $U_i$ and three material dependent parameters --- the density $\rho$, elasticity tensor $c_{ijkl}$ and viscosity tensor $\eta_{ijkl}$ --- and satisfy the acoustic wave equation~\cite{B.A.Auld1973}
\begin{equation}\label{eq:acousticwave}
   - F _ { i } = - \rho \partial _ { t } ^ { 2 } U _ { i } + \sum _ { j k l } \partial _ { j } \left[ c _ { i j k l } + \eta _ { i j k l } \partial _ { t } \right] \partial _ { k } U _ { l }\;.
\end{equation}
In the context of SBS, the driving force $F_i$ results from the electrostriction and radiation pressure \cite{Rakich2010,Wolff2015waveguides} induced by the interfering optical waves. By using the modal description of the pump and signal modes, and the appropriate ansatz given earlier, we find that the force $F_i$ (blue arrows in Fig.~\ref{fig:setup}(b)) from signal and pump pulses has a spectrum (blue shaded region) centred at $\pm(\Omega,~q)=(\omega_p-\omega_S,~k_p-k_S)$, i.e. $F_i(z,t) = f_i(z,t) \exp[i(qz-\Omega t)] + \mathrm{c.c.}$ (note that in Fig.~\ref{fig:setup}(b) we only denote the positive wavenumber components). For resonant SBS, $(\Omega,~q)$ matches the dispersion of a particular acoustic mode, driving the field $\mathbf{U}(\mathbf{r},t) = b(z,t) \tilde{\mathbf{u}}(\mathbf{r}_\perp) \exp[i(qz - \Omega t)] +  \mathrm{c.c.}$. For optical forces with a narrow wavenumber spectrum, as shown in Fig.~\ref{fig:setup}(b), we can separate the positive and negative wavenumber components of the optical forces and acoustic fields --- this is equivalent to a \textit{spatial} RWA. However, if the optical force has a wide frequency spectrum, it may excite acoustic waves propagating in the opposite direction, centred at $\pm(-\Omega,~q)$, which are also modes of the waveguide. Physically, this corresponds to a case where the optical fields interact in a spatially extended region, as to precisely define the acoustic wavenumber, but too quickly to properly define the velocity of the acoustic field driven by a wide spectrum of frequencies. In this picture, the acoustic ansatz should be rewritten to account for the negative velocity waves: $\mathbf{U} = b_+ \tilde{\mathbf{u}}\exp[i(qz - \Omega t)] + b_- \tilde{\mathbf{u}}\exp[i(qz + \Omega t)] +  \mathrm{c.c.}$. However, the choice of envelopes $b_\pm$ is arbitrary, and dictated solely by convenience of the expansion around a particular point in the frequency-wavenumber space. In particular, we can readily define the entire acoustic field by setting $b_-\equiv 0$, and allowing $b_+$ to support the entire acoustic spectrum shown in Fig.~\ref{fig:setup}(b). This however means that the envelope $b_+$ must evolve rapidly in time, as it needs to account for terms oscillating at $-\Omega$. Noting that the positive and negative wavenumber components of the optical force and the acoustic field are decoupled from each other, we can remove the $\mathrm { c.c. }$ terms from both sides of the equation, thereby applying a \emph{spatial} RWA. Furthermore, to retrieve the formulation reminiscent of \eqref{eq:acoustic}, we can follow the derivation from Ref.~\cite{Wolff2015waveguides}, by projecting the above equation onto the mode profile $\mathbf{u}$, and identifying the average power flow over the waveguides' cross-section and the loss term as the inverse dissipation length found in equations (16), (18) and (45) of Ref.~\cite{Wolff2015waveguides}, respectively. As in that work, we identify terms proportional to $\partial_z^2 b_+ +2i q \partial_z b_+$, which for a slowly spatially varying envelope $b_+$ satisfy $|\partial_z^2 b_+| \ll |q \partial_z b_+|$, allowing us to drop the second derivative, and thus embrace the \textit{spatial} SVEA.

We also find terms $\partial_t^2 b_+ - 2i\Omega \partial_t b_+$ which, following our previous observation that the envelope $b_+$ include components oscillating with frequency $-2\Omega$, cannot be simplified through a \textit{temporal} SVEA. Therefore, we finally arrive at the equation for short acoustic pulses driven by SBS:
\begin{equation}\begin{aligned}\label{eq:acequationfinal}
     &\frac{i}{2\Omega v_\mathrm{a}}\partial_t^2 b +  \left(\frac{1}{v_\mathrm{a}}+\frac{i}{\Omega}\alpha\right)\partial_t  b + \partial_z  b + \alpha b 
    \\=& -\frac{i\Omega}{\mathcal{P}_\mathrm{o}}\mathcal{Q}_\mathrm{a} a_{\mathrm{p}}a_{\mathrm{s}}^*\;,
\end{aligned}\end{equation}
where we dropped the index writing $b_+=b$ for convenience. Comparing it to the long-pulse formulation given in Eq.~\ref{eq:acoustic}, we identify the unchanged coupling term, loss and first order derivatives. Our considerations on short pulse Brillouin interactions add a second order time derivative of the envelope and a modified loss term proportional to the first order derivative in time.

Eqs.~(\ref{eq:pump})--(\ref{eq:acoustic}) and~(\ref{eq:acequationfinal}) are nonlinear coupled partial differential equations with no analytical solution. Earlier numerical studies relied upon undepleted pump approximations, stationary or soliton solutions~\cite{Keaton2014,Montes1997}. Here we obtain numerical solutions resolved in space and time for arbitrary initial conditions, coupling parameters and combinations of wavelengths by employing a split-step method~\cite{Bandrauk1993,Feit1982}. The linear time evolution is computed by a Crank-Nicholson method with symmetric differencing in time and space and the nonlinear part by the 4th-order Runge-Kutta algorithm. From the initial complex field vectors of $a_{\mathrm{p}},a_{\mathrm{s}},b,B({t=0,z})$ we calculate the next time step by alternating between linear and nonlinear evolution. We introduce an auxiliary field $B = \partial_t b$ and therefore can write $   \partial_t B = 2(i\Omega + \alpha v_\mathrm{a}) B  + 2i\Omega v_\mathrm{a} \partial_z b  -  2i\Omega v_\mathrm{a} \alpha b - {2\Omega^2 v_\mathrm{a}} \mathcal{Q}_\mathrm{a} a_\mathrm{p}a_\mathrm{s}^*/{\mathcal{P}_\mathrm{o}}$ to obtain first order differential equations from Eq.~(\ref{eq:acequationfinal}), giving a system of four first order equations together with Eqs.~(\ref{eq:pump}) and~(\ref{eq:Stokes}). 

\begin{figure}[!ht]
    \centering
    \includegraphics[width=\linewidth]{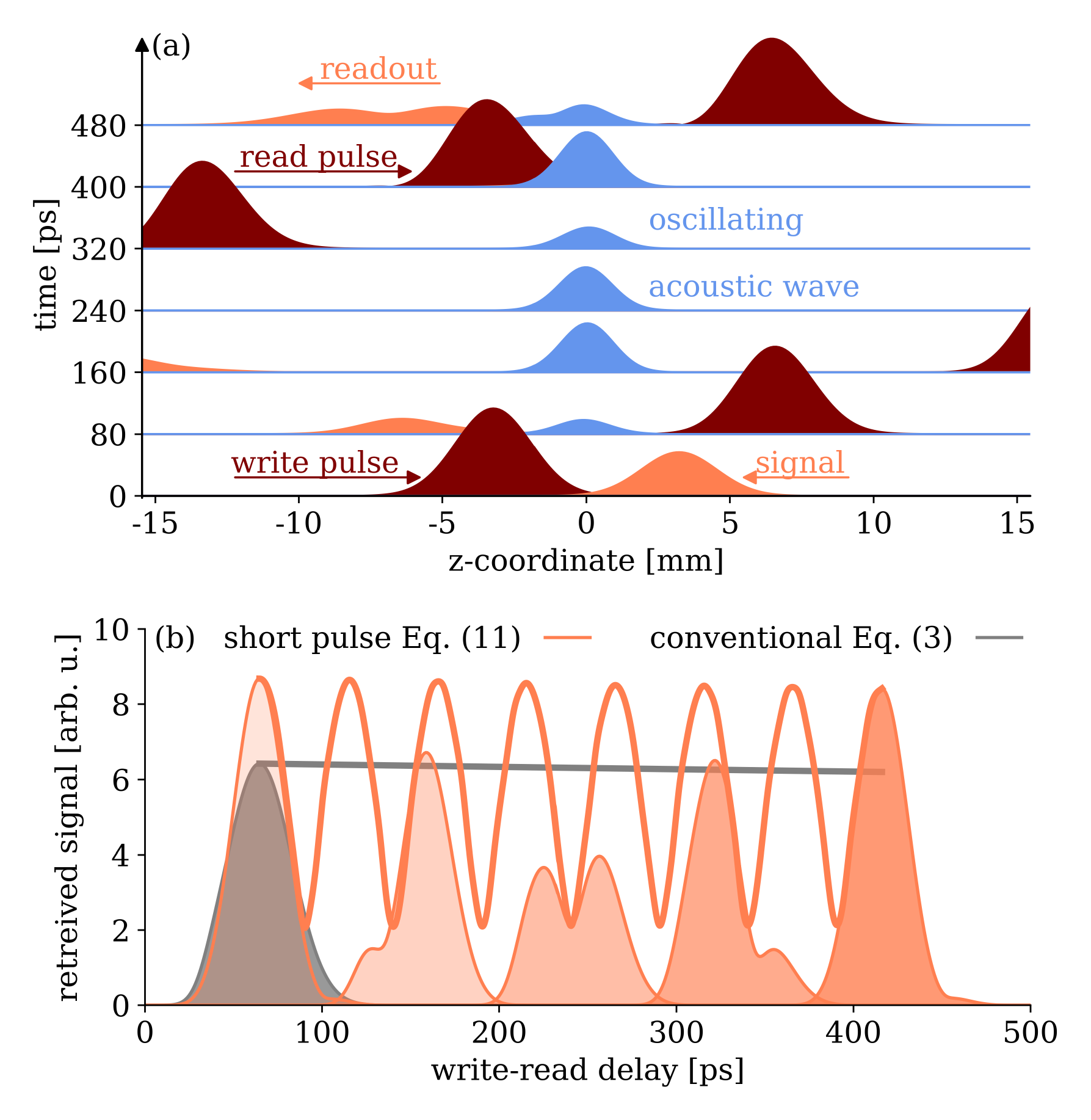}
    \caption{(a): Simulation of SBS storage shows the buildup of an acoustic wave during overlap of the optical signal and write pulses of \SI{50}{\pico\second} length and the subsequent partial retrieval of the signal by a second read pulse after a \SI{400}{\pico\second} delay. (b): The retrieved signal (shaded orange), with examples shown for 60, 150, 240, 330 and \SI{420}{\pico\second} delay time, exhibits modified pulse shapes. The enveloping curve (orange line) oscillates with $2\Omega/2\pi = \SI{20}{\giga\hertz}$. It deviates from the simple slow exponential decay with lifetime $1/\alpha v_\mathrm{a} = \SI{10}{\nano\second}$ exhibited by the readout (grey line) when simulating the conventional Eq.~(\ref{eq:acoustic}).}
    \label{fig:WriteRead}
\end{figure}

The effects of the full modified acoustic Eq.~(\ref{eq:acequationfinal}) describing short acoustic pulses can be observed in the system of SBS-based optical data storage. Here, a strong \emph{write} (pump) pulse interacts with a weaker \emph{signal} pulse and transfers some of its energy into an acoustic pulse of the same shape. A second \emph{read} (pump) pulse inverts this process and partially restores the optical signal (see Fig.~\ref{fig:WriteRead}(a)). All inputs are Gaussian pulses. We trace the resulting optical signal in Fig.~\ref{fig:WriteRead}(b) for the input of two $\tau = \SI{50}{\pico\second}$ optical pulses and an acoustic mode with $\Omega/2\pi = \SI{10}{\giga\hertz}$. The absolute values of five readout signal envelopes ($a_\mathrm{s}$) are marked by the orange shaded areas, where each is shifted in time by the delay time (60, 150, 240, 330 and \SI{420}{\pico\second}) between the write and read pump pulses. One example of the readout resulting from the conventional Eq.~(\ref{eq:acoustic}) for \SI{60}{\pico\second} delay is underlaid in grey as a reference. We calculate 500 more readouts of modified and conventional Eq.~(\ref{eq:acequationfinal}) and Eq.~(\ref{eq:acoustic}), spanning from 60 to \SI{420}{\pico\second}, and draw their enveloping lines in orange and grey, respectively. We omit results for delay times shorter than the pulse widths, as write and read pulse significantly overlap in this case. Both the conventional and short pulse solutions show a slow decay due to the finite acoustic lifetime chosen as \SI{10}{\nano\second}. As $\tau<2\pi/\Omega$ we are in the short pulse regime and the solution to the short pulse equation deviates significantly from the conventional one, exhibiting \SI{20}{\giga\hertz} oscillations. These are a surprising feature of SBS storage with short pulses, as usually the storage efficiency decays over time in accord with the acoustic lifetime. Here, reading out at a later time may result in more peak signal recovered when coinciding with a maximum of the enveloping line. Additionally, the readout signal's shape is modified from the conventional Gaussian pulse, with a modulating two-peak structure visible during the transition between the maxima.

To develop an intuition for the effect of short optical pulses inducing the generation of an acoustic wave, we can simplify the modified acoustic Eq.~(\ref{eq:acequationfinal}) by neglecting the spatial propagation terms (on the basis of the very low group velocity relative to the optical field) and solving it formally as
\begin{align}
    \label{eq:formalb}
    b(t) = &\frac{1}{2\sqrt{(v_\mathrm{a} \alpha)^2-\Omega^2}}   \\ \nonumber
    \bigg[\bigg.&\left. -\int_0^t \tilde{Q}(t') e^{ \left[\left(i\Omega - v_\mathrm{a} \alpha\right)-\sqrt{(v_\mathrm{a} \alpha)^2-\Omega^2}\right] (t-t')/2} dt' \right. \\ \nonumber 
    &+ \bigg. \int_0^t \tilde{Q}(t') e^{ \left[\left(i\Omega - v_\mathrm{a} \alpha\right)+\sqrt{(v_\mathrm{a} \alpha)^2-\Omega^2}\right] (t-t')/2} dt' \bigg],
\end{align}
where $\tilde{Q}(t)=-2\Omega^2 v_\mathrm{a} \mathcal{Q}_\mathrm{a} a_{\mathrm{p}}(t)a_{\mathrm{s}}^*(t)/\mathcal{P}_\mathrm{o}$ describes the optical driving terms. Considering the low velocity of the acoustic wave, it is reasonable to assume this as the physical system of a local ($z=z_0$) acoustic buildup from two counter-propagating optical beams with envelopes $a_{p/s}(t,z=z_0)$. Note, that counter-propagating pulses with width $\tau$ will result in an optical driving term $\tilde{Q}$ with half the width $\tau/2$. For high-mechanical quality acoustic waves $\alpha v_\mathrm{a} \ll \Omega$, we find that the kernel of the second integral oscillates quickly with frequency $2\Omega$. Therefore, unless the optical force $\tilde{Q}$ varies at timescales comparable to $(2\Omega)^{-1}$, this integral averages to $0$. This observation formalises the initial discussion for the applicability of the RWA for the acoustic dynamics. We note, that for low quality acoustic waves $\alpha v_\mathrm{a} > \Omega$ the acoustic wave will quickly decay in overdampened oscillations even for long pulse interactions $\tau>2\pi/\Omega$. These systems are however unlikely to be experimentally viable.

\begin{figure}
\centering
    \includegraphics[width=\linewidth]{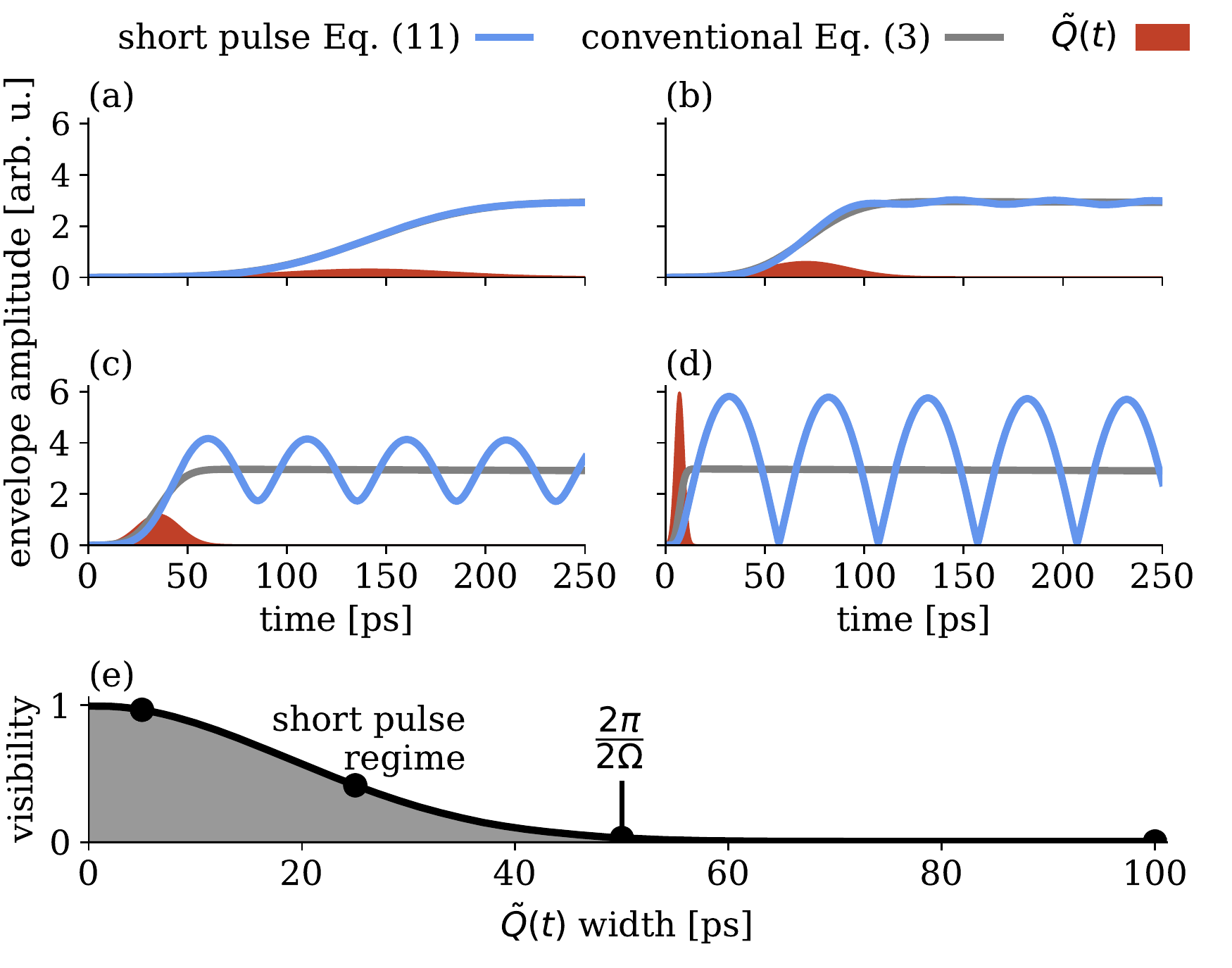}
\caption{Comparing the short (blue) and long (grey) pulse simulations while neglecting spatial dependence reveals their deviation at short time scales. The acoustic wave is driven by two counter-propagating optical pulses resulting in an optical driving term  $\tilde{Q}(t)$ (orange) of width (a) \SI{100}{\pico\second}, (b) \SI{50}{\pico\second}, (c) \SI{25}{\pico\second}, (d) \SI{5}{\pico\second} with an amplitude adjusted to keep the acoustic energy constant. (e): When $\tilde{Q}(t)$ becomes shorter than \SI{50}{\pico\second} (corresponding to \SI{100}{\pico\second} optical pulses), the built up acoustic envelope enters the pulsed regime (shaded) and its dampened harmonic oscillations with $\Omega/2\pi = \SI{20}{\giga\hertz}$ become visible.}
\label{fig:TransitionGain}
\end{figure}

We observe the behaviour of the high quality case in Fig.~\ref{fig:TransitionGain}(a-d), which shows the buildup of the acoustic wave (blue lines) due to the optical driving force $\tilde{Q}(t)$ (red shape) with decreasing pulse widths. The amplitudes of the Gaussian envelopes are scaled up to keep the total pulse energy constant. As we reduce the width to $\Omega^{-1}$ ($\sim\SI{100}{\pico\second}$  for $\Omega/2\pi=10$~GHz), we find oscillations in $b(t)$. We can compare this behaviour to that described by the first order formulation (\eqref{eq:acoustic} with the first spatial derivative removed), yielding $b^{(1)}(t)= i/2\Omega \int_0^t d t' \exp[-v_\mathrm{a} \alpha(t-t')]\tilde{Q}(t')$ shown with grey lines, which diverge from~\eqref{eq:formalb} for short pulse lengths. More quantitatively, we can define the interferometric visibility of the oscillations by the peak difference $[\max_{t>\tau} |b(t)| - \min_{t>\tau} |b(t)|] / [\max_{t>\tau} |b(t)| + \min_{t>\tau} |b(t)|]$ calculated within the first few oscillations. The visibility is shown in Fig.~\ref{fig:TransitionGain}(e), with points marking the cases of Fig.~\ref{fig:TransitionGain}(a-d) and the short pulse regime ($\tau/2<\SI{50}{\pico\second}$) shaded. Finally, in Fig.~\ref{fig:TransitionGain}(d) we consider a response of the acoustic field to the Gaussian optical driving force with \SI{25}{\pico\second}  width (\SI{50}{\pico\second} optical pulses), and find a significant deviation between the predictions of the second ($b$) and first ($b^{(1)}$) order formulations. As this uses the same parameters as Fig.~\ref{fig:WriteRead}(b), we can ascribe the oscillations in the optical readout of short pulsed SBS storage to the oscillations of the intermediate acoustic wave.

In conclusion, we show that the applicability of current descriptions of SBS in waveguides are limited to pulse widths larger than the acoustic period time ($\tau>2\pi/\Omega\approx \SI{100}{\pico\second}$) as they rely on approximating the interacting fields as slowly varying envelopes, and assuming that the direction of the generated acoustic waves in precisely defined. We extend the theoretical description by reevaluating the spatial and temporal RWA and SVEA, and obtain a more suitable equation for dynamics of the acoustic field. Our approach remains valid until spatial oscillations and the optical periods become relevant at around \SI{1}{\femto\second}. The addition of second order terms gives rise to dampened oscillations of the acoustic field when excited by short optical pulses. Our results offer a reference for any Brillouin system when operating in the short pulse regime with the relevant time scales depending on their particular acoustic frequency. 

\section*{Funding}
 Authors acknowledge funding from Australian Research Council (ARC) (Discovery Project DP160101691), the Macquarie University Research Fellowship Scheme (MQRF0001036) and the Max Planck Gesellschaft through an independent Max Planck Research Group.

\bibliography{Bibliography}


\end{document}